\documentclass[a4paper]{jpconf}
\usepackage{graphicx}
\usepackage[skip=2pt,font=scriptsize]{caption}
\usepackage{subcaption}
\usepackage[version=4]{mhchem}
\usepackage{hyperref} 
\hypersetup{
	colorlinks=true,       
	linkcolor=blue,        
	citecolor=blue,        
	filecolor=magenta,     
	urlcolor=blue         
}
\usepackage{isotope}
\begin{document}
\title{Lifetimes and moments measurements to investigate the structure of midheavy nuclei}
\author{T.J.~Mertzimekis, A.~Khaliel, D.~Papaioannou$^1$, G.~Zagoraios, A.~Zyriliou}

\address{Department of Physics, National and Kapodistrian University of Athens, Zografou Campus, GR--15784, Athens, Greece}

\ead{tmertzi@phys.uoa.gr}

\begin{abstract}

Mid--heavy nuclei offer unique opportunities to study the collective and
single--particle aspects of nuclear structure. This mass regime is a dynamic
area where protons and neutrons generally occupy different orbitals, giving
rise to complex structures with a wide variety of shapes, shape evolution
and shape coexistence. To that end, measurements of nuclear lifetimes and
electromagnetic moments ($\mu$,$Q$) can be invaluable. Recent experimental
activities of the NuSTRAP group in Athens have focused on $\gamma$--spectroscopy
studies employing the RoSPHERE array in Magurele, Romania. In recent
studies~\cite{2016_Bucher,2017_Bucher}, the neutron--rich \isotope[144-146]{Ba}
isotopes have exhibited octupole degrees of freedom. Similar questions exist
for the lighter \isotope[140]{Ba} isotope, which is located at the onset of
octupole collectivity. In addition, understanding the structure of heavier,
even--even hafnium isotopes requires more data regarding shape coexistence
and shape evolution. Preliminary results on lifetimes in this area pave the
way to understand dynamical phenomena prior to studying more exotic species
in the future.

\end{abstract}


\section{Introduction}

The entire nuclear landscape is formed due to the existence of two competing forces:
the nuclear attraction and the Coulomb repulsion. As general as this statement might
sound, it is the interplay of those two forces, which can stabilize or destabilize
groups of nucleons so as to form  the $\approx3500$ isotopes known today. The lack of a
rigorous mathematical description of the aggregate nuclear field is a major drawback in
describing the effects of the interplay on nuclear structure and has been at the
forefront of nuclear studies for several decades. As one moves through different areas
of the nuclear chart, the number of degrees of freedom that need be included in theoretical
modeling grows exponentially with mass. Experimental data are necessary to improve
theoretical description.

The mid--heavy nuclei ($A\sim130-180$) belong to a part of the nuclear chart where
protons and neutrons occupy different major shells. Compared to medium--weight
nuclei (e.g. in the vicinity of Kr isotopes), where both protons and neutrons
occupy similar orbitals in the $fp$ shell giving rise to interesting
phenomena~\cite{2003_Mertzimekis}, mid--heavy nuclei feature different Fermi energies
for protons and neutrons and a larger variety of spins. The advent of radioactive beams
together with the rise of interest for nuclear processes in stellar environments have
recently put the region of mid--heavy nuclei into scrutiny. The structure of exotic
species in this mass regime is largely unknown, where spectroscopic data are scarce
even for isotopes a few nucleons away from the valley of stability, thus making any
information on lifetimes, transition rates, wavefunctions or even production cross
sections quite important for understanding the occurrence of dynamical phenomena.

One special way to study the nuclear structure in mid--heavy nuclei is by means
of the nuclear electromagnetic moments, in particular the magnetic dipole moment,
$\mu$ and the electric quadrupole moment, $Q$. In the quantum picture both these
observables are one--body operators, so they can scale up with the number of nucleons.
The magnetic dipole moment has the unique ability to provide the wavefunction of a
nuclear state in terms of its single--particle orbitals occupied by its constituents,
protons and neutrons. As protons and neutrons have markedly different magnetic properties
originating from their distinct spin contributions, the structure of any nuclear state
is reflected on its magnetic moment. On the other hand, the electric quadrupole moment
is directly related to the shape of the nucleus and its deformation. $Q$ can be also
related to the reduced matrix elements B(E2). That relation is model--dependent, however
it provides the means to extract the shape evolution in a particular nucleus by measuring
the lifetimes of states, which are connected to B(E2) values, see for example~\cite{1995_Nilsson}.
In addition, the intriguing phenomenon of shape coexistence is reportedly expected
in mid--heavy nuclei providing additional motivation for detailed studies~\cite{2011_Heyde}.

The present work reports on preliminary experimental results focusing on the nuclear
structure of \isotope[140]{Ba} with poorly known spectroscopic data, other than those of
the ground--state, as well as \isotope[180]{Hf}, where the structure of a considerable number
of observed non-band levels is unknown. In the following sections, the motivation to study
each of these isotopes, the experimental approach and the first results of the ongoing
analysis are presented in some detail. Lessons learnt and future research directions are
also described in the concluding section.

\subsection*{The case of \isotope[140][56]{Ba}}

\begin{figure}[ht]
\centering
\includegraphics[width=0.95\textwidth]{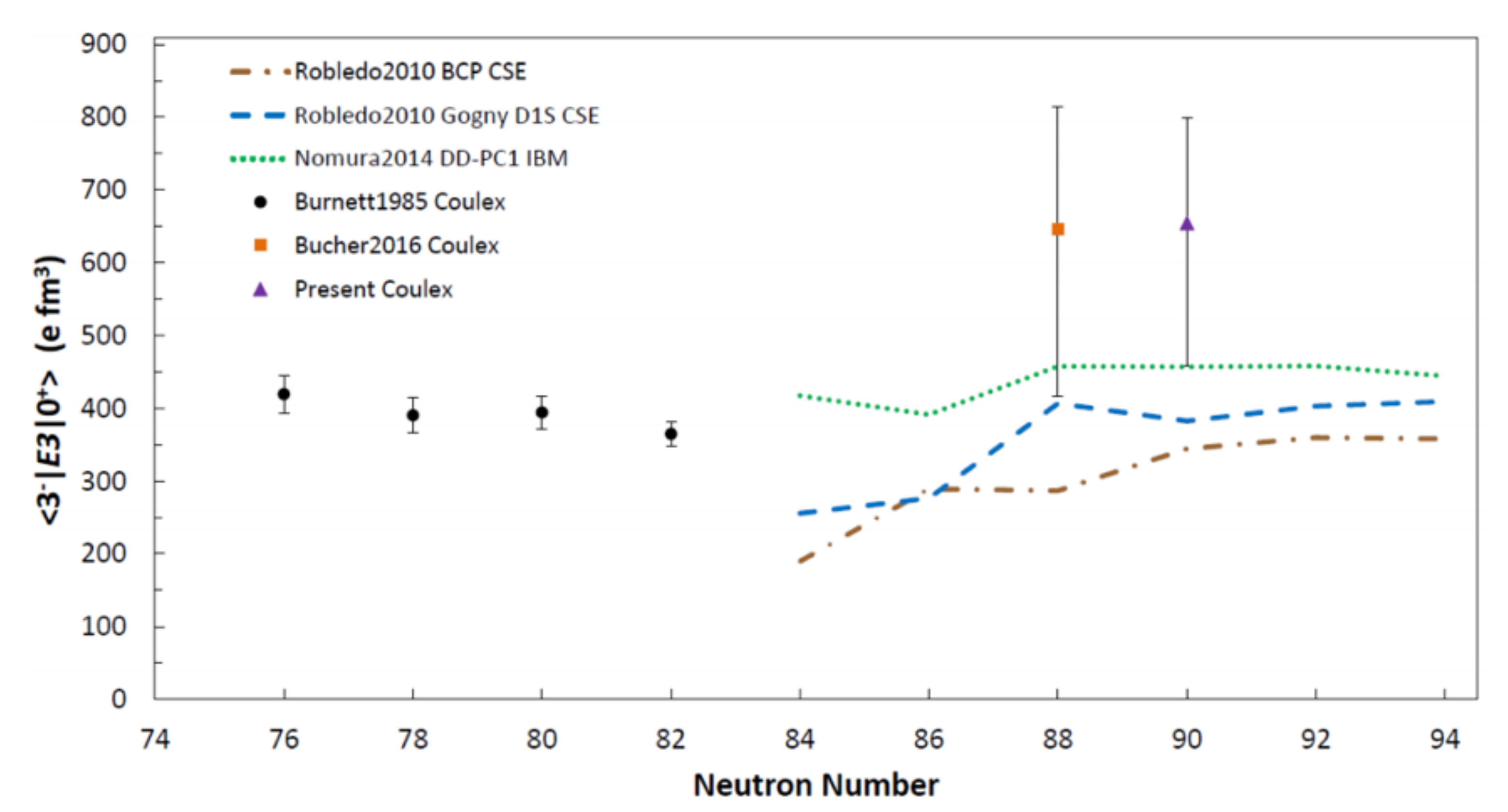}
\caption{%
Experimental $B(E3)$ data for the barium isotopic chain. No data exist for \isotope[140,142]{Ba}.
The figure has been adopted from~\cite{2016_Bucher,2017_Bucher}.
}
\label{fig:barium}
\end{figure}

The \isotope[140]{Ba} nucleus is an unstable nucleus ($t_{1/2}=12.75$~d) located at the
onset of octupole correlations. Two neighbouring isotopes, the neutron--rich \isotope[144,146]{Ba}
isotopes, have been recently studied experimentally in terms of their $B(E3)$
values~\cite{2016_Bucher,2017_Bucher}, using radioactive beams and Coulomb excitation.
Despite some large uncertainties (see Fig.~\ref{fig:barium}) the respective $B(E3)$ values
were found to deviate significantly from any theoretical prediction. As a consequence, the
case of \isotope[140]{Ba} lays important questions on the onset and evolution of octupole
correlations, as well as on the assessment of the degree of collectivity in the barium
isotopic chain as a function of the neutron number.

\begin{figure}[ht]
\centering
\includegraphics[width=0.75\textwidth]{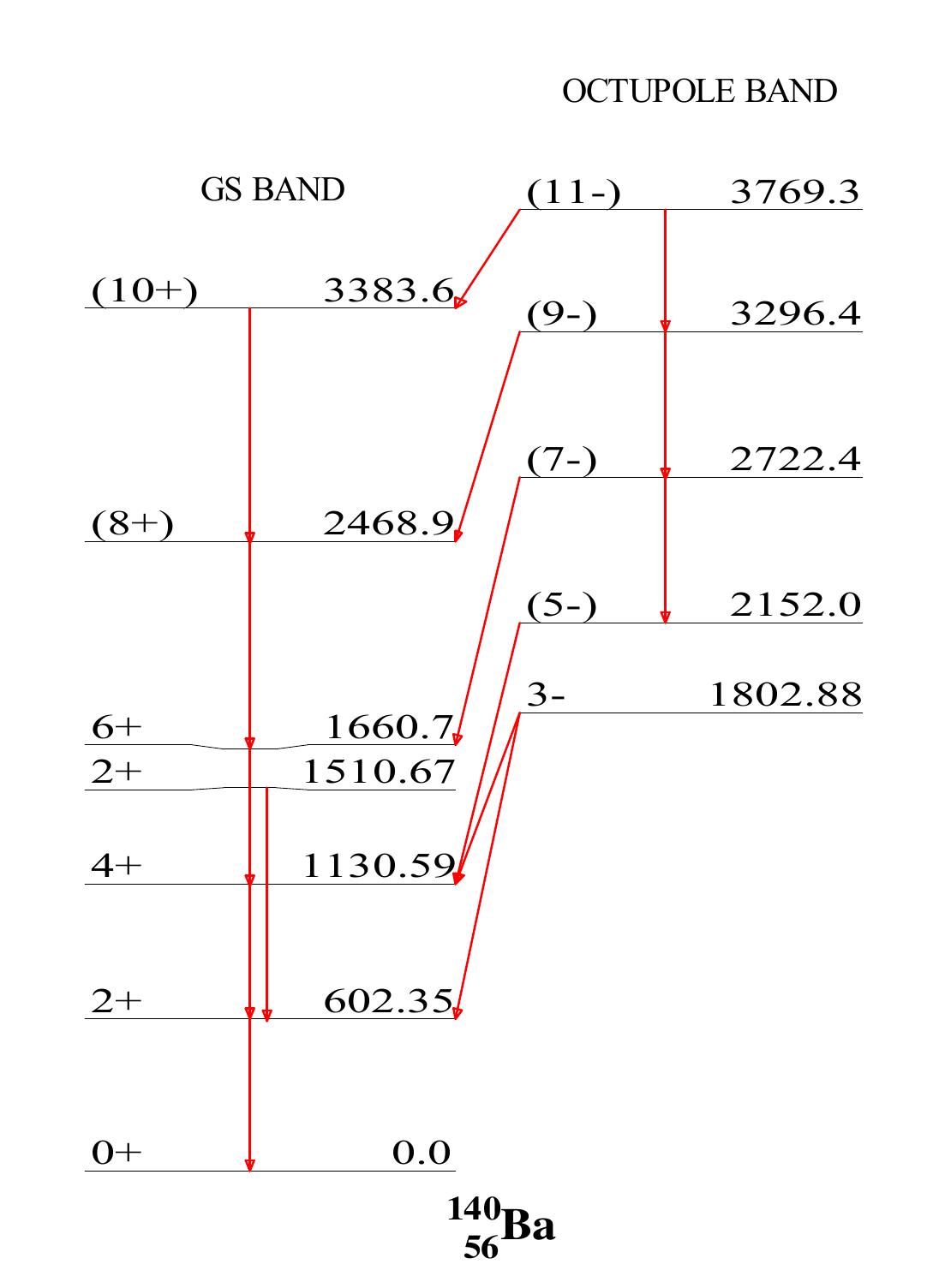}
\caption{%
A partial level scheme of \isotope[140]{Ba} (source~\cite{IAEA}).
}
\label{fig:levelBa}
\end{figure}

Lifetimes of the lower--lying states in \isotope[140]{Ba} are generally unknown, with the sole
exception of the first $2^+$ state~\cite{2012_Bauer}. It is clear that even lower/upper limits
of lifetime measurements can have an impact on understanding the nuclear structure in
\isotope[140]{Ba}. This is particularly relevant for the negative parity band (Fig.~\ref{fig:levelBa}),
which suggests the occurrence of octupole degrees of freedom. From a technical point of view,
there is a certain degree of difficulty in populating the side--band states via Coulomb excitation,
which favors E2 excitations strongly. Thus, a 2n--transfer reaction was chosen instead to populate
states during the experiment. The advantage from using such a reaction is two--fold in the present
case, as \isotope[140]{Ba} can be populated directly from a stable isotope target (\isotope[138]{Ba}).
Chemically, barium is very reactive with air, thus making target--manufacturing a real challenge,
if one needs to preserve target integrity and stoichiometry.

In addition, cross section data related to the production of \isotope[140]{Ba} via the
2n--transfer reaction at energies near and below the Coulomb barrier do not exist for the
case of bombardment with heavy ions, such as \isotope[18]{O}. Such experimental data can be
proven useful for populating neutron--rich nuclei near the neutron dripline.

\subsection*{The case of \isotope[180][72]{Hf}}

\isotope[180]{Hf} is a stable nucleus with an extensive level scheme known from earlier
studies~\cite{NNDC}. The ground state band features a typical rotational spectrum and lifetimes
of its states are known up to spin 12$^+$. An interesting feature of the nuclear structure is
a negative parity band and a few K--isomers, such as the K=4 8$^-$ at 1141.6~MeV with halflife
of 5.53~h, that acts as the bandhead of another rotational band. Questions about its structure
arise from the fact that several non--band members exist with no data present in literature,
and no information on shapes and deformations. The recently developed proxy-SU(3)
model~\cite{2018_Assimakis,2018_Martinou} predicts \isotope[180]{Hf} to be a nucleus with shape
coexistence, but this requires experimental confirmation. Lifetimes and quadrupole moments
can provide useful information to that end, therefore a dedicated study is highly desired.

\section{Experimental Details}

Two separate three--day long experiments were carried out at the 9~MV Tandem Accelerator
Laboratory IFIN--HH in Magurele, Romania using the the RoSPHERE array.

In the first experiment, the 2n--transfer reaction
\isotope[138]{Ba}(\isotope[18]{O},\isotope[16]{O})\isotope[140]{Ba}
was employed at four energies just below the Coulomb barrier (61, 63, 65 and 67~MeV).
The target was made with a ``sandwich--like'' structure after an elaborate procedure
that placed 2~mg/cm$^2$ of barium between a thick Au backing of 4.88~mg/cm$^2$, and
an evaporated 0.5~mg/cm$^2$ Au layer at the front. RoSPHERE was loaded with 15 HPGe detectors
and 10 LaBr$_3$ scintillators (the latter were not used).

The second experiment used a thick 5~mg/cm$^2$ \isotope[nat]{Ta} foil to produce \isotope[180]{Hf}
via the charge--pickup reaction \isotope[181]{Ta}(\isotope[11]{B},\isotope[12]{C})\isotope[180]{Hf}
at a beam energy of 47~MeV. Following de--excitation, emitted $\gamma$ rays were recorded by 25 HPGe distributed over
(five) 5 rings.

\section{Results and Discussion}

\begin{figure}[ht]
\centering
\includegraphics[width=0.98\textwidth]{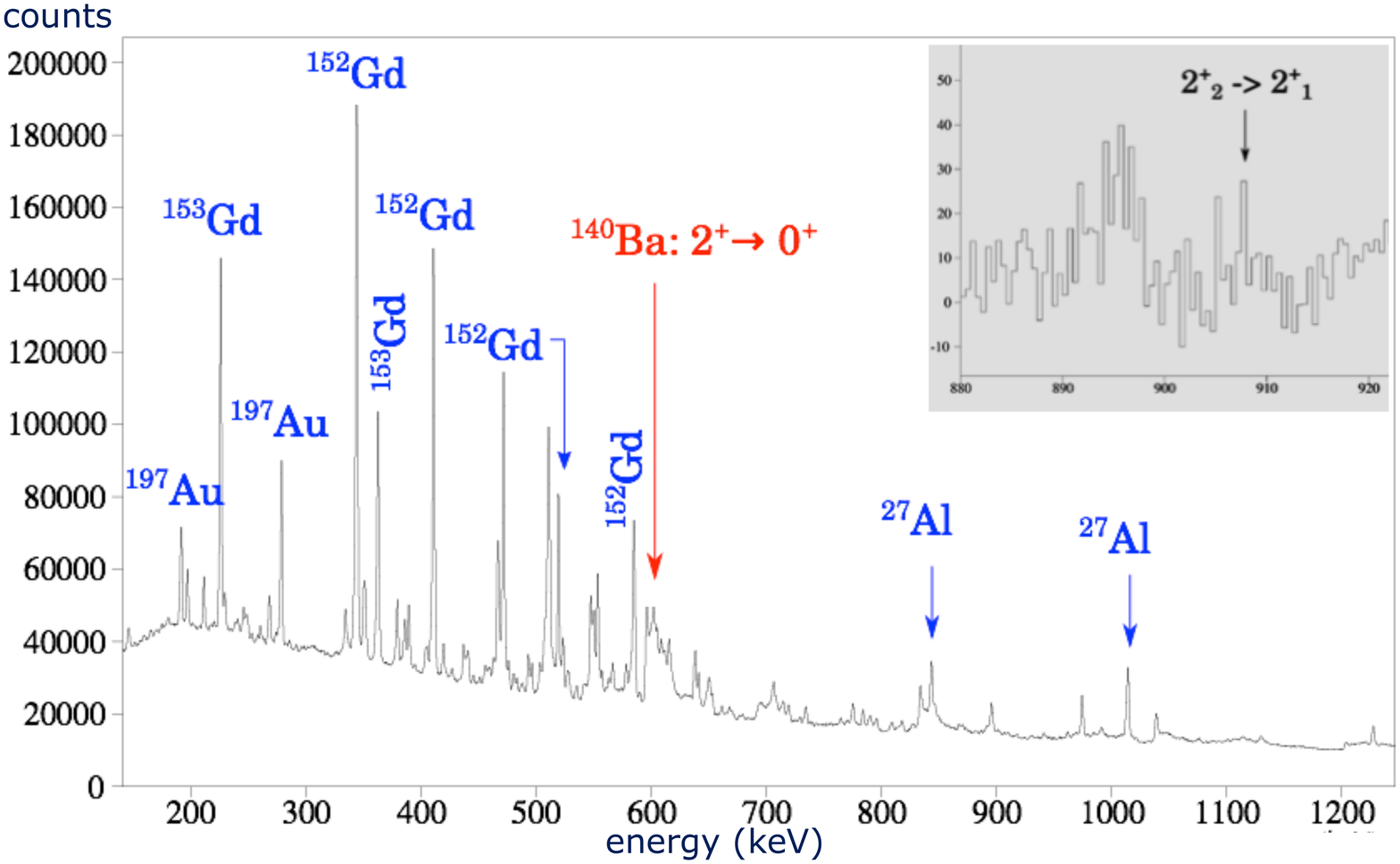}
\caption{%
A total statistics spectrum of \isotope[140]{Ba} in the forward ring. The $2_1^+\rightarrow 0_{gs}^+$ transition is marked. The inset marks the $2_2^+\rightarrow 2_{1}^+$ transition.
}
\label{fig:specBa}
\end{figure}

In the first experiment, the beam energies of 61 and 63~MeV resulted in very low statistics
for the 2n--transfer reaction, while the maximum energy of 67~MeV was sufficient to trigger
strong, competing reaction channels that dominated the recorded spectra. The 65~MeV energy was
a convenient choice that avoided the aforementioned drawbacks. A typical spectrum of
\isotope[140]{Ba} is shown in Figure~\ref{fig:specBa}. At this energy setting, the levels
in the ground--state band were populated up to 8$^+$. Unfortunately, levels in the side
band, in particular those that are potentially the best indicators for studying quadrupole
and octupole collectivity (see $0_2^+$ and $3_1^-$ in Figure~\ref{fig:levelBa}) have not been
populated sufficiently during the total 21 hours of data collection.

An attempt to measure lifetimes in \isotope[140]{Ba} in the g.s. band was performed with the
Doppler--Shift Attenuation Method (DSAM). Forward--backward lineshapes were searched in the
37--143$^\circ$ ring spectra, without much success. Given the limitations of DSAM, a lower
lifetime limit can be safely set to $\approx 1$~ps.

Further analysis of the \isotope[140]{Ba} data focused on the production cross sections.
Relative cross sections of the 2n--transfer reaction have been measured with respect to the 
competing fusion--evaporation channel \isotope[138]{Ba}(\isotope[18]{O},4n)\isotope[154]{Gd}.
The results of this part of the analysis are reported elsewhere~\cite{2019_Khaliel}.

\begin{figure}[ht]
\centering
\includegraphics[width=0.98\textwidth]{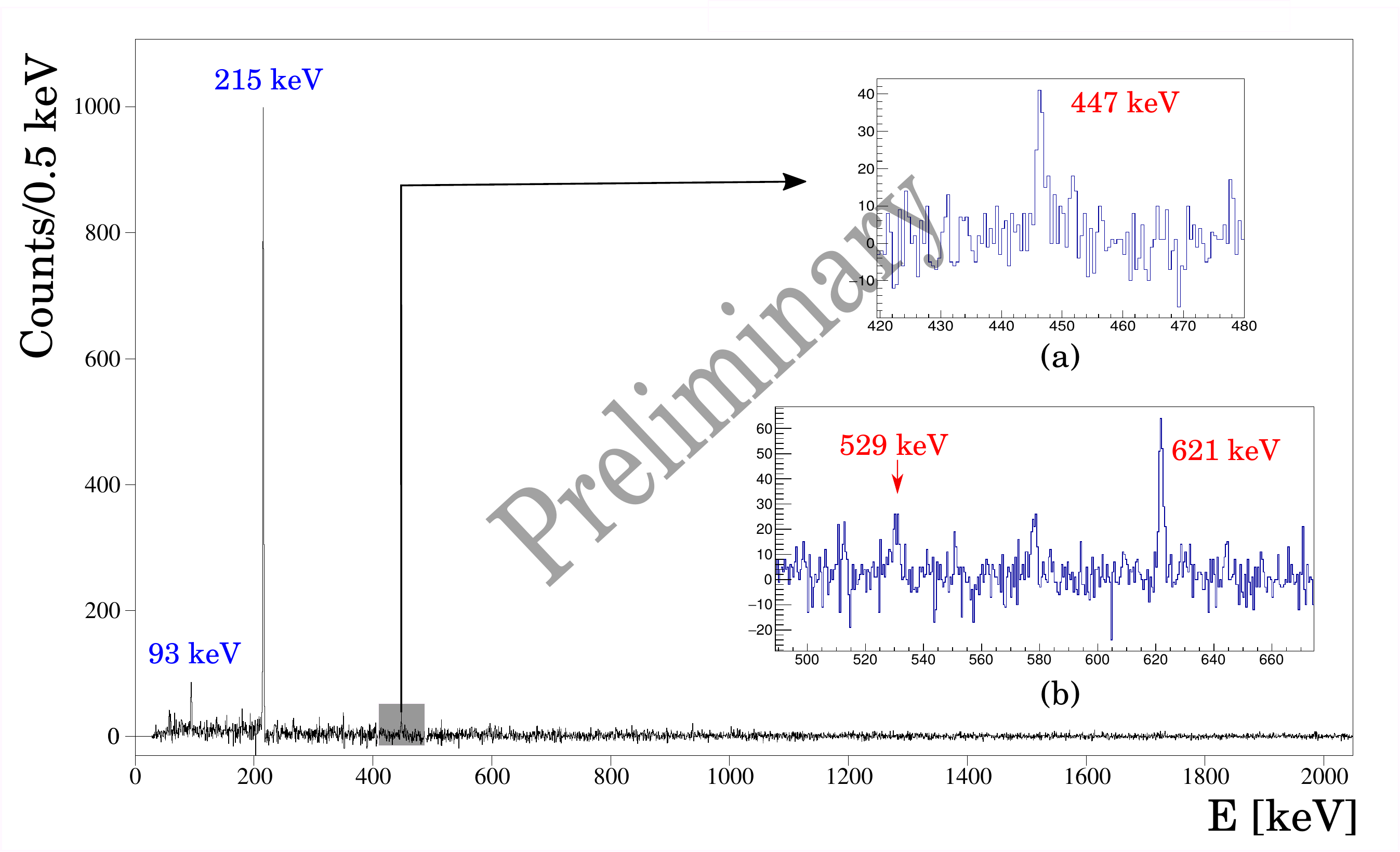}
\caption{%
A spectrum produced after gating on the 1066~keV transition, $(4^-)\rightarrow 4^+_1$. In inset (a)
the events coming in coincidence with the 447~keV transition depopulating the non--band 1821~keV
state (3$^-$) in \isotope[180]{Hf} are shown. In inset (b), part of the coincidence spectrum
gated on the transitions 1198 ($3^+_1\rightarrow 2^+_1$) and 1200~keV ($2^+_3 \rightarrow g.s.$)
shows the 621 and 529~keV transitions, which also depopulate the 1821~keV state. The structure
and the lifetime of this state is currently unknown.
}
%
\label{fig:specHf}
\end{figure}

In the second experiment, approximately three (3) full days of experimental data were collected.
The focus on this experiment was on both the ground state band, which has unknown lifetimes, but
also on the negative parity band that has an isomeric state as a bandhead ($J^\pi=4^-$, $t_{1/2}=0.57~\mu\textrm{s}$). This part of the analysis is ongoing and has already offered interesting results
on the \isotope[180]{Hf} nuclear structure. A focus on $\gamma-\gamma$ angular correlations has
already shown sufficient statistics to suggest or re-examine spin/parity assignments and offers
the opportunity to study the unknown structure of non--band levels depopulated from $\gamma$
transitions as shown in Figure~\ref{fig:specHf}, where the 1821 non--band state is depopulated
by a $\gamma$ transition of 447 keV.

\section{Conclusions and Future Work}

The present work reports on two different test experiments focusing on mid--heavy nuclei:
the lighter, unstable \isotope[140]{Ba} and the heavier, but stable, \isotope[180]{Hf},
respectively. In the latter experiment, the suspected occurrence of octupole collectivity
could not be confirmed due to very low statistics. In the particular case, the 2n--transfer
reaction cannot be used for the population of states that could provide lifetimes and reduced
matrix elements relevant to octupole degrees of freedom, despite it has been successful in
other cases, such as in \isotope[66]{Ni}~\cite{2017_Leoni}. As barium is a difficult target
to make and use in a stable beam experiment (e.g. using a plunger device to measure lifetimes),
a plausible solution is to consider a radioactive beam experiment, likely in inverse kinematics.
A lower lifetime limit of $\tau >1$~ps was determined based on the limitations of DSAM, but
further work is necessary to that direction.

For the case of \isotope[180]{Hf}, the analysis is ongoing and seems very promising. In the
near future, the focus will be on angular correlations of states, search for short--lived
states that could be measured with DSAM and the scrutiny of the nature of states that do not
belong in a particular band and may provide evidence for large deformation or isomerism.
The charge--pickup reaction seems to produce sufficient statistics, so as one could
potentially use it in a plunger experiment for measuring lifetimes with low uncertainty,
thus providing data to study the evolution of transition matrix elements and shapes in
\isotope[180]{Hf}.


\section*{Acknowledgements}
This research work is supported by the Hellenic Foundation for Research
and Innovation (HFRI) and the General Secretariat for Research and Technology
(GSRT) under the HFRI PhD Fellowship Grant (GA. No. 74117/2017). TJM acknowledges
partial financial support by the Bulgarian National Science Fund (BNSF) under 
Contract No. KP-06-N28/6.

\end{document}